\documentclass[aps,prl,twocolumn,superscriptaddress,footnotebib]{revtex4-1}
\usepackage{graphicx}
\usepackage{amssymb,amsmath}
\usepackage{bm}
\usepackage{dcolumn}
\usepackage{subfigure}
\usepackage{float}
\usepackage[OT1]{fontenc} 
\usepackage{url}
\usepackage{slashed}
\usepackage{color}
\usepackage{verbatim}
\usepackage{txfonts}
\usepackage[driverfallback=dvipdfm]{hyperref}
\hypersetup{pdfpagemode=FullScreen,colorlinks=true,breaklinks,urlcolor=blue,linkcolor=blue,citecolor=blue}
\usepackage{natbib}
\usepackage{verbatim}

\usepackage{subfigure}
\usepackage{comment}
\usepackage{hyperref}
\usepackage{amssymb}
\usepackage{bm}
\usepackage{graphicx}
\usepackage{amsmath,amssymb}
\usepackage{tikz,fp}
\usepackage{tikz-cd}
\usetikzlibrary{arrows}
\usetikzlibrary{intersections}
\usetikzlibrary{shapes.geometric}
\usetikzlibrary{decorations.pathmorphing, patterns,shapes,fixedpointarithmetic}
\usetikzlibrary{decorations.markings}

\pgfdeclarepatternformonly{south west lines}{\pgfqpoint{-0pt}{-0pt}}{\pgfqpoint{3pt}{3pt}}{\pgfqpoint{3pt}{3pt}}{
	\pgfsetlinewidth{0.4pt}
	\pgfpathmoveto{\pgfqpoint{0pt}{0pt}}
	\pgfpathlineto{\pgfqpoint{3pt}{3pt}}
	\pgfpathmoveto{\pgfqpoint{2.8pt}{-.2pt}}
	\pgfpathlineto{\pgfqpoint{3.2pt}{.2pt}}
	\pgfpathmoveto{\pgfqpoint{-.2pt}{2.8pt}}
	\pgfpathlineto{\pgfqpoint{.2pt}{3.2pt}}
	\pgfusepath{stroke}}

\tikzset{
	mid arrow/.style={postaction={decorate,decoration={
				markings,
				mark=at position .575 with {\arrow{stealth}}
	}}},
	near arrow/.style={postaction={decorate,decoration={
				markings,
				mark=at position .275 with {\arrow{stealth}}
	}}},
	far arrow/.style={postaction={decorate,decoration={
				markings,
				mark=at position .800 with {\arrow{stealth}}
	}}},
	snake arrow/.style={fixed point arithmetic, decorate, decoration={snake,amplitude=2pt, segment length=11pt},postaction={decoration={markings,mark=at position 0.625 with {\arrow{stealth}}},decorate}},
}

\begin{document}
	
	\title{Violation and Revival of Kramers' Degeneracy in Open Quantum Systems}
	
	\author{Pengfei Zhang}
	\affiliation{Walter Burke Institute for Theoretical Physics, California Institute of Technology, Pasadena, CA 91125, USA}
	\affiliation{Institute for Quantum Information and Matter, California Institute of Technology, Pasadena, CA 91125, USA}

	\author{Yu Chen}
	\thanks{ychen@gscaep.ac.cn}
	\affiliation{Graduate School of China Academy of Engineering Physics, Beijing, 100193, China}

	\date{\today}
	
	\begin{abstract}
        Kramers' theorem ensures double degeneracy in the energy spectrum of a time-reversal symmetric fermionic system with half-integer total spin. Here we are now trying to go beyond the closed system and discuss Kramers' degeneracy in open systems out of equilibrium. In this letter, we prove that the Kramers' degeneracy in interacting fermionic systems is equivalent to the degeneracy in the spectra of different spins together with the vanishing of the inter-spin spectrum. We find the violation of Kramers' degeneracy in time-reversal symmetric open quantum systems is locked with whether the system reaches thermal equilibrium. After a sudden coupling to an environment in a time-reversal symmetry preserving way, the Kramers doublet experiences an energy splitting at a short time and then a recovery process. We verified the violation and revival of Kramers' degeneracy in a concrete model of interacting fermions and we find Kramers' degeneracy is restored after the local thermalization time. By contrast, for time-reversal symmetry $\tilde{\cal T}$ with $\tilde{\cal T}^2=1$, we find although there is a violation and revival of spectral degeneracy for different spins, the inter-spin spectral function is always nonzero. We also prove that the degeneracy in spectral function protected by unitary symmetry can be maintained always. 
	\end{abstract}
	
	\maketitle
{\color{blue}\emph{Introduction.}} --  
 Kramers' degeneracy theorem tells us for fermionic systems with half-integer total spin where time-reversal symmetry (TRS) is presented, all energy levels are doubly degenerate~\cite{kramers, wigner}. This theorem plays a vital role in the quantum spin Hall effect~\cite{Kane,Shoucheng} as well as in the stability of the superconducting phase with disorder~\cite{Anderson}. 
 
It is natural to expect that Kramers' theorem applies to thermal equilibrium systems since the grand canonical distribution is only related to the hamiltonian where the double degeneracy is presented. This can be proved straightforwardly in equilibrium systems~\cite{lieu2021kramers}. On the other hand, according to McGeely and Cooper, TRS is not stable in open quantum systems even through the interactions between the system and the bath bares TRS~\cite{cooper2020}. The original statement is for a pure state as a superposition of Kramers' states can not maintain its coherence after coupling to the environment. Its consequences for symmetry protected topological states are later studied\cite{knapp2020,cai2021}. Even more surprising, by non-Hermitian linear response theory~\cite{pan2020nhlt}, it is discovered that after coupling to the environment in a time-reversal symmetry preserving way, the Kramers' degeneracy is lifted and two helical topological edge states in topological insulator will mix with each other~\cite{deng2020stability}. As a grand canonical ensemble can be understood as the steady-state of the open quantum system, these two results seem to be contradicted with each other.  

To clarify the seeming paradox, in this work we go beyond the perturbation theory and study the whole dynamical process, focusing on the Kramers' degeneracy for a time-reversal symmetric interacting fermion system with a TRS preserving interaction between the system and the bath. We find that the Kramers' degeneracy is locked with thermal equilibrium. We find a violation of Kramers' degeneracy after a sudden coupling to an environment and the violation enlarges with time until a certain time scale when it shrinks. The Kramers' degeneracy revives as the system gradually reaches new thermal equilibrium. Although the Kramers' degeneracy comes back, the irreversibility shows itself in the spectrum change in the final state.
  	\begin{figure}[tb]
		\centering
		\includegraphics[width=0.95\linewidth]{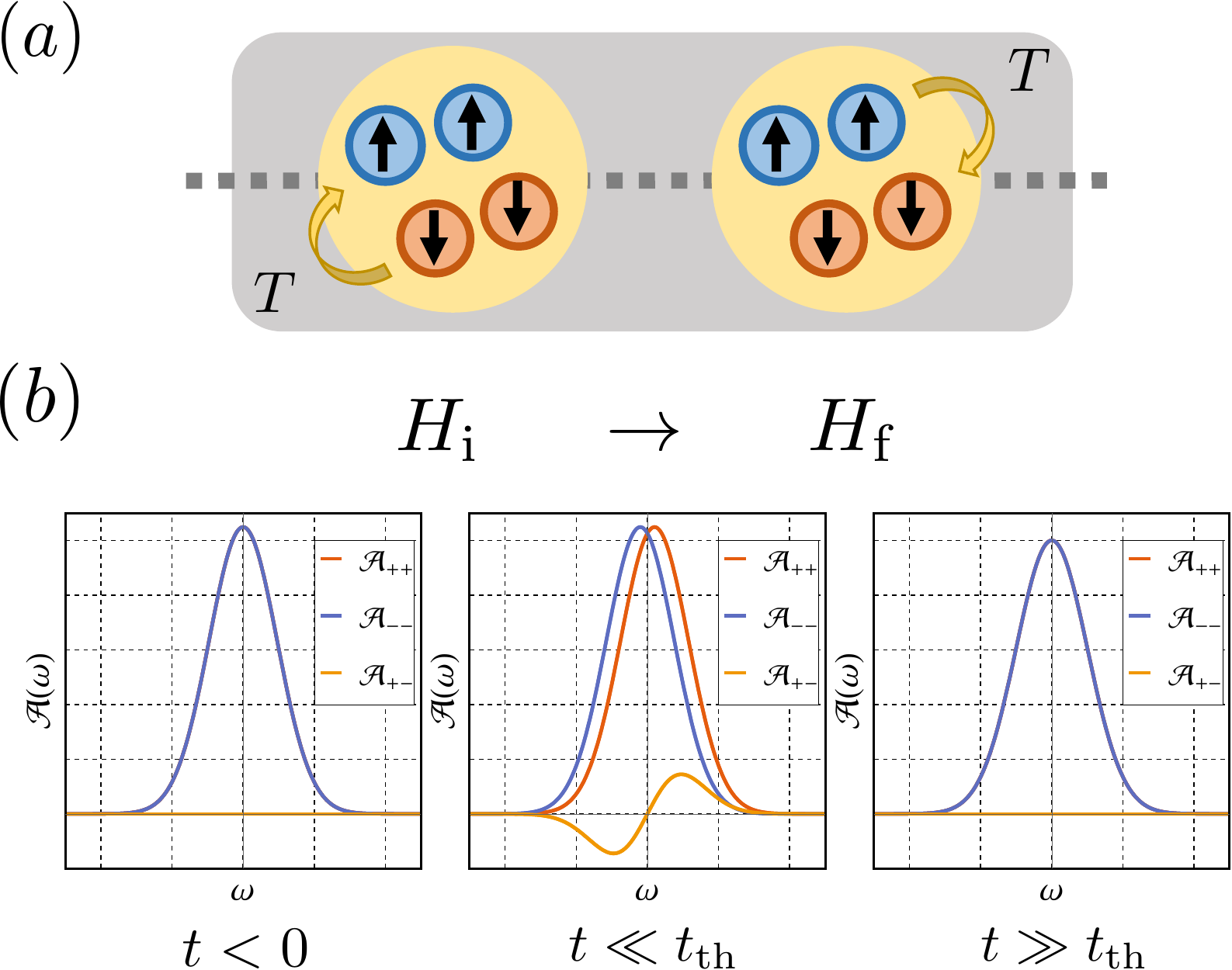}
		\caption{ Schematics of (a). Spin-$1/2$ fermionic models (yellow) coupled to an external bath (gray). The time-reversal symmetry interchanges spin $\uparrow$ and $\downarrow$, with additional phases. (b). The spectral function during a quantum quench, with a breaking and restoring of Kramers' Degeneracy with characteristic time scale $t_{\text{th}}$. Here we take the real-part of $\mathcal{A}_{+-}$.}
		\label{fig:schemticas}
	\end{figure}

In the following, we first establish a general relation between Kramers' degeneracy and single-particle spectral functions. Then with the help of solvable interacting models, we can calculate the quench dynamics of the spectral function and the distribution function with a sudden coupling to a bath. By checking the time scale of distribution function reaching thermal equilibrium and Kramers' degeneracy signal in spectral functions, we can draw our conclusion. We also discussed how spectral functions evolute for system-to-bath interaction baring time-reversal symmetry $\tilde{\cal T}$ with $\tilde{\cal T}^2=1$ and unitary symmetry ${\cal S}$ as a comparison. We stress that our conclusion applies to general interacting fermion systems and is not restricted by the Markovian approximation. By this study, we find Kramers' degeneracy only emerges in thermal equilibrium systems, which implies TRS is only a good symmetry in equilibrium system and the breaking extent of Kramers' degeneracy can be set as a measure of the extent the system being away from equilibrium.

  {\color{blue}\emph{General Theory.}} --  We consider a quantum many-body system with spin-$1/2$ fermions $\hat{c}_{j,\sigma}$, where $\sigma=\pm$ labels spin states, and $j$ is the mode index. Under the time-reversal transformation, the fermionic annihilation operator $\hat{c}_{j,\sigma}$ satisfies $\hat{\cal T}\hat{c}_{j,\sigma}\hat{\cal T}^{-1}=\sigma \hat c_{j,-\sigma}$ and thus $\hat{\cal T}^2=(-1)^{N_s}$. Here $N_s$ is the fermion number in the system. We prepare the system in thermal equilibruim of initial Hamiltonian $\hat{H}_{i}$ as $\hat{\rho}_0=e^{-\beta_{0}\hat{H}_{i}}/Z$, with $Z=\text{tr}~[e^{-\beta_0 \hat{H}_{i}}]$. We assume $\hat{H}_{i}$ is the system hamiltonian with time-reversal symmetry satisfying $\hat{\cal T} \hat{H}_{i} \hat{\cal T}^{-1}=\hat{H}_{i}$. Kramers’ theorem states that the eigenstate of $\hat{H}_{i}$ with odd $N_s$ should have a pairwise degeneracy~\cite{bernevig2013topological}: Given any eigenstate $|\psi_1\rangle$, satisfying $\hat{H}_{i}|\psi_1\rangle=E_0|\psi_1\rangle$, one can show that for $|\psi_2\rangle\equiv \hat{\cal T}|\psi_1\rangle$, it satisfies $\hat{H}_{i}|\psi_2\rangle=E_0|\psi_2\rangle$ and $\langle \psi_1|\psi_2\rangle=0$. As a comparison, $\hat{H}_{i}|\psi_2\rangle=E_0|\psi_2\rangle$, without $\langle \psi_1|\psi_2\rangle=0$, also works in the even $N_s$ subspace and is not sufficient for proving the existence of the degeneracy.

However, for general interacting quantum systems, it is hard to probe a specific eigenstate experimentally. It is Green's function of local operators that can be measured by various experimental protocols~\cite{altland2010condensed}, which in general give the spectrum and distribution function information of the quasi-particles. Here we are trying to discuss the Kramers' degeneracy in open quantum systems, where even no eigenstates of the system are well defined. Therefore we have to introduce Green's function form of the Kramers' degeneracy, which also works in the open quantum systems. First, we introduce real-time Green's functions $G^{\gtrless}(t,t')$, which are defined as
  \begin{equation}
  \begin{aligned}
iG^>_{\sigma\sigma'}(t,t')&\equiv \text{tr}~\left[\hat{\rho}_0 \hat{c}_{j,\sigma}(t)\hat{c}^\dagger_{j,\sigma'}(t')\right],\\
-iG^<_{\sigma\sigma'}(t,t')&\equiv \text{tr}~\left[\hat{\rho}_0 \hat{c}^\dagger_{j,\sigma'}(t')\hat{c}_{j,\sigma}(t)\right],
  \end{aligned}
  \end{equation}
where $\hat{c}_{j,\sigma}(t)=e^{i\hat{H}_{i}t}\hat{c}_{j,\sigma}^{}e^{-i\hat{H}_{i}t}$ is fermion annihilation operator in Heisenberg picture. Here ${\rm tr}$ is over the Hilbert space of the system.
  Then it is straighforward to show that there is an analogy of the Kramers' theorem $G^{\gtrless}_{++}(t,t')=G^{\gtrless}_{--}(t,t')$ and $G^{\gtrless}_{+-}(t,t')=G^{\gtrless}_{-+}(t,t')=0$ in thermal equilibrium~\cite{lieu2021kramers}. Here, $G^{\gtrless}_{+-}(t,t')=G^{\gtrless}_{-+}(t,t')=0$, as an analogy of $\langle \psi_1|\psi_2\rangle=0$, is a signature of having $\hat{\cal T}^2=(-1)^{N_s}$. Using their relation to the spectral function $\mathcal{A}=i(G^>-G^<)/2\pi$, we have $\mathcal{A}_{++}=\mathcal{A}_{--}$ and $\mathcal{A}_{+-}=\mathcal{A}_{-+}=0$ in thermal equilibrium. 

  In this work, we are interested in understanding the Kramers' degeneracy in the quench dynamics from the Green's function perspective. At $t=0$, we change the Hamiltonian from $\hat{H}_{i}$ to $\hat{H}_f$, which also satisfies $\hat{\cal T} \hat{H}_f \hat{\cal T}^{-1}=\hat{H}_f$. In particular, we couple the original system to an additional bath by taking $\hat{H}_f=\hat{H}_{i}+\hat{V}$, where $\hat{V}=\sum_{j\ell}V_{j\ell}\hat{\cal O}_j\hat{\xi}_\ell$, satisfying $\hat{\cal T}\hat{\cal O}_j\hat{\cal T}^{-1}=\hat{\cal O}_j$ and $\hat{\cal T}\hat{\xi}_j\hat{\cal T}^{-1}=\hat{\xi}_j$. Operators $\hat{\cal O}_j$ and $\hat{\xi}_j$ are time-reversal symmetric operators in the system and the environment. Nevertheless, we also comment on the quantum quench of isolated quantum systems. Previous studies show the Kramers’ degeneracy is broken when the system is coupled to the bath~\cite{cooper2020,deng2020stability}. More generally, using the time-reversal transformation, one can show that for $t,t'>0$ we have
  \begin{equation}\label{bath}
  \begin{aligned}
i(G^>_{++}-G^>_{--}) =\left<\text{tr}\left[\Delta\hat \rho(t,t')~e^{i\hat{H}_f(t-t')}c_{j,-}e^{-i\hat{H}_f(t-t')}c^\dagger_{j,-}\right]\right>_B,&\\
-2iG^>_{\sigma,-\sigma} =\left<\text{tr}\left[\Delta\hat \rho(t,t')~e^{i\hat{H}_f(t-t')}c_{i,\sigma} e^{-i\hat{H}_f(t-t')}c^\dagger_{i,-\sigma}\right]\right>_B.&
  \end{aligned}
  \end{equation}
  Here $\Delta\hat \rho(t,t')\equiv\hat \rho(-t)-\hat \rho(t')=e^{iH_ft}\hat \rho_0e^{-i\hat{H}_ft}-e^{-i\hat{H}_ft'}\hat \rho_0e^{i\hat{H}_ft'}$ and $\left<...\right>_B=\text{tr}_B~(\hat \rho_B...)$ is the quantum expectation on the bath density matrix, which is assumed to be thermal with inverse temperature $\beta_{\rm B}$. Since the bath contains a much larger degree of freedom, we assume that it does not evolve when coupled to the small system. Similar relations hold for $G^<_{\sigma\sigma'}$ and thus $\mathcal{A}_{\sigma\sigma'}$.

  In the short time limit, we have $\Delta\hat \rho(t,t')\approx i(t+t')[\hat{H}_f,\hat{\rho}_0]+\frac{t'^2-t^2}{2}[\hat{H}_f,[\hat{H}_f,\hat{\rho}_0]]+O(t^3)\neq 0$. For $\left<\hat{\xi}_j\right>=0$, the leading-order contribution is from $[\hat{H},[\hat{H},\hat{\rho}_0]]$. Considering that $\hat{\rho}_0=e^{-\beta_0\hat{H}_{i}}$, therefore $[\hat{H}_f,\hat{\rho}_0]=[\hat{V},\hat{\rho}_0]=\sum_{j\ell}V_{j\ell}\hat{\xi}_\ell[\hat{\cal O}_j,\hat{\rho}_0]$. Then we can see, the time-reversal symmetry of $\hat{\cal O}_j$ does not ensure $[\hat{\cal O}_j,\hat{\rho}_0]=0$, therefore in general $[\hat{H},[\hat{H},\hat{\rho}_0]]\neq 0$.  As a result, $G^{\gtrless}_{++}\neq G^{\gtrless}_{--}$, $G^{\gtrless}_{\sigma,-\sigma}\neq 0$ and the Kramers' degeneracy is broken. This generalizes the previous analysis using Markovian baths~\cite{deng2020stability}. As a comparison, for unitary symmetries interchanging $+$ and $-$, one finds a similar relation for $i(G^>_{++}-G^>_{--})$, but with $\Delta\hat{\rho}(t,t')=\hat{\rho}(t')-\hat{\rho}(t')=0$. It means the degeneracy protected by unitary symmetry is stable in dynamical evolutions. We also notice that the condition that $\Delta\hat{\rho}(t,t')\neq 0$ is very general and has a more natural understanding than the extended Schur Lemma argument given previously.
  
 On the other hand, for $t,t'\gg t_\text{th}$, where $t_\text{th}$ is the thermalization time, we expect the total system to thermalize and $\hat \rho_{\text{tot}}(-t)\approx \hat \rho_\text{tot}(t')\approx e^{-\beta_f (\hat{H}_f+\hat{H}_B)}/{\rm tr}_B[\text{tr}[e^{-\beta_f (\hat{H}_f+\hat{H}_B)}]]$ under simple probes. Here $\hat H_B$ is the Hamiltonian of the bath. Note that this final density matrix is a thermal ensemble of the coupled system, where the system-bath interaction can be important. Consequently, we find $G^{\gtrless}_{++}= G^{\gtrless}_{--}$, $G^{\gtrless}_{\sigma\sigma'}= 0$, and the Kramers' Degeneracy is restored. For a large system-bath coupling, the characteristic time scale $t_\text{th}$ resembles its counterpart in isolated quantum systems, where $t_\text{th}\sim \beta_f$ for strongly interacting models and $t_\text{th}\sim 1/\Gamma$ with quasi-particle decay rate $\Gamma$ for weakly interacting models~\cite{hartnoll2018holographic,stanford2016many,zhang2020obstacle}. 

  Before turning to concrete examples, we add a few comments. Firstly, the violation and revival of Kramers' Degeneracy also exists in isolated quantum systems, where $\hat{H}_f$ is only different from $\hat{H}_i$ by certain parameters. In this case, \eqref{bath} still works, without the average over bath density matrix. In the long-time limit, although the fine-grained density matrix $\rho(t)$ may differs from the thermal density matrix since the unitary evolution preserves the total entropy, we expect local thermalization $\hat \rho(-t)\sim \hat \rho(t')\sim e^{-\beta_f \hat{H}_f}/\text{tr}[e^{-\beta_f \hat{H}_f}]$~\cite{Deutsch_1991,Srednicki_1994}. Here $\sim$ means the equivalence under measurement of local operators. 

  Secondly, it is helpful to compare the above results with anti-unitary symmetry with $\tilde{\cal T}^2=1$ or unitary symmetry $\mathcal{S}$. We choose the single-particle transformation to be $\mathcal{\tilde{T}}=\sigma_x K$ or $\mathcal{S}=i\sigma_y$. Here $K$ is the operator for taking the complex conjugate. In both cases, the symmetry imposes $G_{++}=G_{--}$ in thermal equilibrium, while generally $G_{+-}\neq 0$. When coupled to the bath, for anti-unitary symmetry $\tilde{\cal T}$, the $G_{++}=G_{--}$ firstly breaks and then gets restored. While for $S$, the $G_{++}=G_{--}$ is always preserved during the evolution.

  {\color{blue}\emph{Concrete Model.}} --  
We now verify our predictions in interacting fermions using a concrete solvable model. Generally, the simulation of quench dynamics of chaotic quantum systems is hampered by the exponential growth of the Hilbert space dimension. Here, we overcome this difficulty by constructing a time-reversal invariant SYK model~\cite{Sachdev_Ye,kitaev2015simple,maldacena2016remarks,davison2017thermoelectric,Gu:2019jub} by coupling different complex SYK sites~\cite{banerjee2017solvable,song2017strongly,chen2017competition,zhang2017dispersive,jian2017solvable}, which is solvable in the large-$N$ expansion. The initial Hamiltonian $H_i$ reads
\begin{equation}
\begin{aligned}
H_i=\sum_{j\tau\sigma\tilde \tau\tilde\sigma}\hat c^\dagger_{j\tau\sigma}h^{\tau\sigma}_{\ \tilde \tau\tilde\sigma} \hat c^{}_{j\tilde \tau\tilde\sigma}+\sum_{\tau\sigma\{j_a\}}\frac{J^{\tau\sigma}_{j_1j_2j_3j_4}}{4}\hat c^\dagger_{j_1\tau\sigma}\hat c^\dagger_{j_2\tau\sigma}\hat c^{}_{j_3\tau\sigma}\hat c^{}_{j_4\tau\sigma}.
\end{aligned}
\end{equation}
Here $j/j_a=1,2,...N$ labels different fermion flavors, $\sigma$ is a spin index and $\tau=\pm$ is an additional pseudo-spin index. $J^{\tau\sigma}_{j_1j_2j_3j_4}$ describes intraspecies random interaction between fermions, which satisfies independent Gaussian distribution with 
\begin{equation}\label{sysJ}
\left<J^{\tau\sigma}_{j_1j_2j_3j_4}\right>=0,\ \ \ \ \ \ \ \ \left<\left|J^{\tau\sigma}_{j_1j_2j_3j_4}\right|^2\right>={2J^2}/{N^3}.
\end{equation}
Under the time-reversal transformation, we find $J^{\tau\sigma}_{j_1j_2j_3j_4}\rightarrow (J^{\tau-\sigma}_{j_1j_2j_3j_4})^*$. As a result, the interaction term is time-reversal invariant after ensemble average. The single-particle Hamiltonian $h^{\tau\sigma}_{\ \tilde \tau\tilde\sigma}$ is diagonal in the flavor space. Imposing the single-particle time-reversal invariance with $\mathcal{T}=i\sigma_y K$, we find generally
\begin{equation}
\begin{aligned}
h~=~&\mu~I\otimes I+K_x \tau_x\otimes I+K_z \tau_z\otimes I\\
&+J_x \tau_y\otimes \sigma_x+J_y \tau_y\otimes \sigma_y+J_z \tau_y\otimes \sigma_z,
\end{aligned}
\end{equation}
where $\mu$ is the chemical potential, $K_x$, $K_z$, $J_{x,y,z}$ are real parameters.
Non-trivial terms correspond to celebrated $\gamma$ matrices wildly used in both condensed matter \cite{bernevig2013topological} and high-energy physics \cite{peskin1995quantum}. As a result, the single-particle eigenstates show pairwise degeneracy at energy $\mu\pm \sqrt{K_x^2+K_y^2+J_x^2+J_y^2+J_z^2}$, consistent with Kramers' theorem.

We consider the quench by coupling the system to an external bath at $t=0$, with system-bath coupling 
\begin{equation}
H_{\text{SB}}(t)=\theta(t)\sum_{j_1j_2b_1b_2}V_{j_1j_1b_1b_2}\left(\sum_{\tau\sigma\tilde \tau\tilde\sigma}c^\dagger_{j_1\tau\sigma}\tilde{h}^{\tau\sigma}_{\ \tilde \tau\tilde\sigma} c^{}_{j_2\tilde \tau\tilde\sigma}\right)\psi^\dagger_{b_1}\psi^{}_{b_2}.
\end{equation}
Here, to be concretes, we choose the bath to be an additional SYK model with $M\gg N$ fermion modes ($b_1,b_2=1,2,...,M$)~\cite{chen2017tunable,zhang2019evaporation,almheiri2019universal,chen2020replica}. This corresponds to 
$
H_{\text{B}}=\sum{J^\text{B}_{b_1b_2b_3b_4}}\psi^\dagger_{b_1}\psi^\dagger_{b_2}\psi_{b_3}^{}\psi_{b_4}^{}/4.
$
We further choose the distribution of $J^\text{B}_{b_1b_2 b_3b_4}$ takes the similar form as \eqref{sysJ}, with $N$ replaced by $M$. Generalizations to other bath models are straightforward. The coupling strength satisfies
\begin{equation}
\left<V_{j_1j_2b_1b_2}\right>=0,\ \ \ \ \ \ \ \ \left<\left|V_{j_1j_2b_1b_2}\right|^2\right>={V^2}/{NM^2}.
\end{equation}
This guarantees that $H_{\text{SB}}$ does not affect the evolution of the bath~\cite{chen2017tunable,zhang2019evaporation,almheiri2019universal,chen2020replica}, consistent with our previous assumption. We also choose the $\tilde{h}$ to take general form 
\begin{equation}
\begin{aligned}
\tilde{h}~=~&\tilde\mu~I\otimes I+\tilde K_x \tau_x\otimes I+\tilde K_z \tau_z\otimes I\\
&+\tilde J_x \tau_y\otimes \sigma_x+\tilde J_y \tau_y\otimes \sigma_y+\tilde J_z \tau_y\otimes \sigma_z,
\end{aligned}
\end{equation}
where $\tilde{\mu}$, $\tilde{K}_x$, $\tilde{K}_y$, $\tilde{J}_{x,y,z}$ are independent parameters.
The form of $\tilde{h}$ ensures the ensemble of couplings are also invariant under the time-reversal symmetry $\hat{\cal T}$. Here we have extended $\hat{\cal T}$ to the full system by defining $\hat{\cal T}\hat{\psi}^{}_b \hat{\cal T}^{-1}=\hat{\psi}_b$. 

\begin{figure}[tb]
		\centering		
\begin{tikzpicture}[scale=1.3, baseline={([yshift=-4pt]current bounding box.center)}]
\filldraw (-20pt,0pt) circle (1pt);
\filldraw (20pt,0pt) circle (1pt);
\draw[thick,mid arrow] (-20pt,0pt) .. controls (-8pt,-14pt) and (8pt,-14pt) .. (20pt,0pt);
\draw[thick,mid arrow] (20pt,0pt) .. controls (8pt,14pt) and (-8pt,14pt) .. (-20pt,0pt);
\draw[thick,mid arrow] (20pt,0pt) -- (-20pt,0pt);
\draw[thick,mid arrow] (-20pt,0pt) -- (-35pt,0pt);
\draw[thick,mid arrow] (35pt,0pt) -- (20pt,0pt);
\node at (23pt,10pt) {\scriptsize $J$};
\node at (-23pt,10pt) {\scriptsize $J$};
\draw[thick, dotted] (20pt,0pt) .. controls (8pt,24pt) and (-8pt,24pt) .. (-20pt,0pt);
\end{tikzpicture}\ \ \ \ \ \ \ \ \ \ \ 
\begin{tikzpicture}[scale=1.3, baseline={([yshift=-4pt]current bounding box.center)}]
\filldraw (-20pt,0pt) circle (1pt);
\filldraw (20pt,0pt) circle (1pt);
\draw[dashed,thick,mid arrow] (-20pt,0pt) .. controls (-8pt,-14pt) and (8pt,-14pt) .. (20pt,0pt);
\draw[thick,mid arrow] (20pt,0pt) .. controls (8pt,14pt) and (-8pt,14pt) .. (-20pt,0pt);
\draw[dashed,thick,mid arrow] (20pt,0pt) -- (-20pt,0pt);
\draw[thick,mid arrow] (-20pt,0pt) -- (-35pt,0pt);
\draw[thick,mid arrow] (35pt,0pt) -- (20pt,0pt);
\node at (23pt,10pt) {\scriptsize $V\tilde{h}$};
\node at (-23pt,10pt) {\scriptsize $V\tilde{h}$};
\draw[thick, dotted] (20pt,0pt) .. controls (8pt,24pt) and (-8pt,24pt) .. (-20pt,0pt);
\end{tikzpicture}
		\caption{ The contribution to the self-energy $\Sigma^\gtrless(t,t')$ in the large-$N$ limit. Here the solid lines represent the Green's function of fermions in the system and the dashed lines represent the bath Green's function. The dotted lines represent the disorder average.} 
		\label{fig:melons}
	\end{figure}
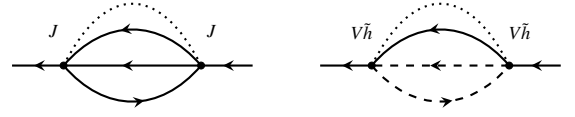

In the large-$N$ limit, the Green's functions $G^{\gtrless}$ of SYK-like models satisfies the Schwinger-Dyson equation on the Keldysh contour, and the quench dynamics can be simulated by solving corresponding integral equations. Explicitly, we have
\begin{equation}\label{num1}
\begin{aligned}
(i\partial_t-h)\circ G^{\gtrless}&=\Sigma^\text{R}\circ G^{\gtrless}+\Sigma^\gtrless \circ G^{\text{A}},\\
G^{\gtrless}\circ(i\partial_{t}-h)&=G^\text{R}\circ \Sigma^{\gtrless}+G^\gtrless \circ \Sigma^{\text{A}}.
\end{aligned}
\end{equation}
Here $\circ$ includes the convolution in real-time, as well as multiplication in $\sigma$ and $\tau$ space. The self-energy is given by melon diagrams shown in Fig. \ref{fig:melons}, which leads to

\begin{equation}\label{num2}
\begin{aligned}
\Sigma^\gtrless_{\tau\sigma,\tau'\sigma'}(t,t')=&J^2\delta_{\sigma\sigma'}\delta_{\tau\tau'}G^\gtrless_{\tau\sigma,\tau\sigma}(t,t')^2G^\lessgtr_{\tau\sigma,\tau\sigma}(t',t)\\
&+V^2\chi^\gtrless_{\text{B}}(t,t')(\tilde{h}G^\gtrless\tilde{h})_{\tau\sigma,\tau'\sigma'}(t,t')\theta(t)\theta(t').
\end{aligned}
\end{equation}
Here $\chi^\gtrless_{\text{B}}(t,t')=G^\gtrless_\psi(t,t')G^\lessgtr_\psi(t',t)$ is the bath correlation function. The retarded/advanced Green's functions are related to $G^\gtrless$ by $G^\text{R}(t_1,t_2)=\theta(t_{12})(G^>(t_1,t_2)-G^<(t_1,t_2))$ and $G^\text{A}(t_1,t_2)=\theta(t_{21})(G^<(t_1,t_2)-G^>(t_1,t_2))$. Similar relations holds for self-energies. Using these relations, \eqref{num1} and \eqref{num2} become closed. The numerical approach for solving \eqref{num1} and \eqref{num2} with discretized time has been well explained in previous works~\cite{zhang2019evaporation,eberlein2017quantum,kuhlenkamp2020periodically,zhou2021disconnecting}.

	\begin{figure*}[tb]
		\centering
		\includegraphics[width=1\linewidth]{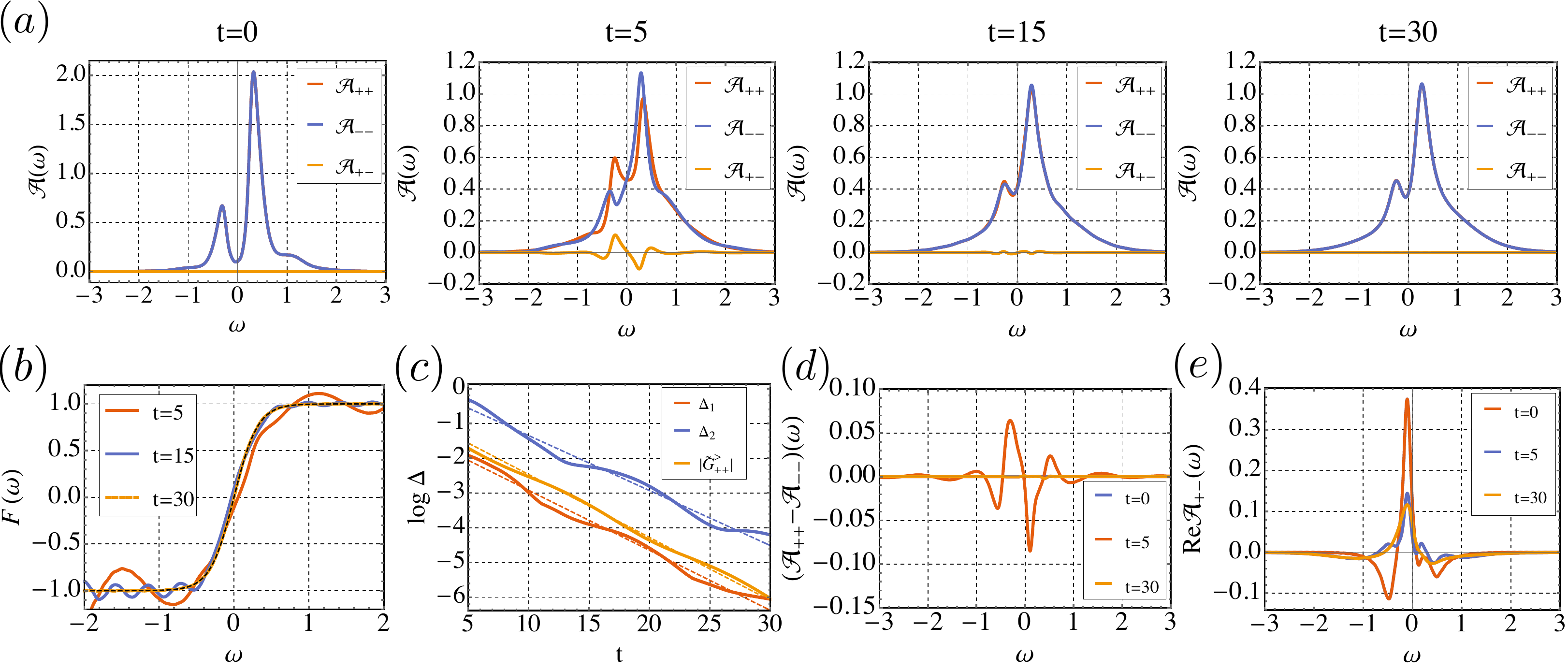}
		\caption{Numerical results for the time-reversal invariant SYK model coupled to bath. We choose $\beta_{\text{i}}J=\beta_{\text{f}}J=6$, $V=J$, and we set $\beta=2\pi$ as the unit of time. We further set $(\mu,K_x,K_z,J_x,J_y, J_z)/J=(0,0.1,0.2,0.3,0.15,0.2)$ and $(\mu,\tilde K_x,\tilde K_z,\tilde J_x,\tilde J_y,\tilde J_z)/J=(0,0.4,0.4,0.6,0.36,0.2)$. (a). The spectral function $\mathcal{A}_{++}$, $\mathcal{A}_{--}$ and $\text{Re}~\mathcal{A}_{+-}$ as a function of evolution time $t$. There is a breaking and restoring of Kramers' Degeneracy. (b). The quantum distribution $F(\omega)$ at different time $t$. The system almost thermalizes at $t=30$. The black dashed curve corresponds to a plot of $\tanh \pi \omega$. (c). The log-plot of $\Delta_1$ or $\Delta_2$ as a function of time $t$, which characterize the breaking of the Kramers' Degeneracy. The dashed lines correspond to the results of the linear fittings. As a comparison, we also plot the evolution of spectral $\mathcal{A}$ for a model~\cite{ST} with symmetry $\tilde{\cal T}$ in (d-e). } 
		\label{fig:num}
	\end{figure*}

  {\color{blue}\emph{Numerical Results.}} --  
  We now present numerical results of the quench dynamics. We choose $\beta_i=\beta_f$, $V=J$, and arbitrarily chosen parameters in $h$ and $\tilde{h}$. Given the real-time Green's function $G^\gtrless(t,t')$, we define the temporal Green's function $\tilde G^\gtrless(t_r,t)$ at time $t$ by 
  \begin{equation}
  \tilde G^\gtrless(t_r,t)\equiv\begin{array}{cc}
  \left\{ 
    \begin{array}{cc}
      G^\gtrless(t+t_r,t)\ \ \ & t_r\leq 0, \\
      G^\gtrless(t,t-t_r)\ \ \ & t_r> 0.
    \end{array}\right.
\end{array}
  \end{equation}
 This definition preserves the causality of the unitary evolution. Here $\tilde{G}^\gtrless$ and $G^\gtrless$ are in matrix form, and the sub-indices are omitted. We define the Fourier transform with respect to $t_r$. The temporal spectral function $\mathcal{A}(\omega,t)$ then reads
\begin{equation}
\mathcal{A}(\omega,t)=\frac{i}{2\pi}\int dt_r~e^{i\omega t_r}( \tilde G^>(t_r,t)-\tilde G^<(t_r,t)).
\end{equation}
In numerics, we focus on the first site with $\tau=+$ and drop the corresponding pseudo-spin indices for conciseness. The results for $\mathcal{A}_{++}(\omega,t)$, $\mathcal{A}_{--}(\omega,t)$, and $\text{Re}~\mathcal{A}_{+-}(\omega,t)$ are shown in Fig. \ref{fig:num} (a). Before the quench, the system is in thermal equilibrium and the Kramers' theorem ensures $\mathcal{A}_{++}(\omega,0)=\mathcal{A}_{--}(\omega,0)$ and $\mathcal{A}_{+-}(\omega,0)=0$. After we couple the system to the bath ($t>0$), the degeneracy is lifted. As an example, we find a large discrepancy between $\mathcal{A}_{++}$ and $\mathcal{A}_{--}$, as well as a non-vanishing $\mathcal{A}_{+-}$ at $t=5$. When the time $t$ becomes longer, $\mathcal{A}_{++}-\mathcal{A}_{--}$ and $\mathcal{A}_{+-}$ decays, and becomes almost invisible at $t=30$.

Our previous analysis shows the revival of Kramers' degeneracy happens when the system arrives at equilibrium with the bath. In a quantum many-body system, the local thermalization can be diagnosed by quantum distribution function $F(\omega,t)$ at time $t$. It can then be defined as 
\begin{equation}
F(\omega,t)\mathcal{A}(\omega,t)=\frac{i}{2\pi}\int dt_r~e^{i\omega t_r}( \tilde G^>(t_r,t)+\tilde G^<(t_r,t)).
\end{equation}
In thermal equilibrium, we have $F(\omega)=1-2n_F(\omega)=\tanh (\beta\omega/2)$, with Fermi-Dirac distribution function $n_F(\omega)$. We plot $F(\omega,t)$ for different $t$ in Fig. \ref{fig:num} (b). Shortly after the quench, $F(\omega,t)$ significantly deviates from the $\tanh (\beta\omega/2)$ and the system is far from equilibrium. At longer time $t=15$, $F(\omega,t)$ approaches the $\tanh (\beta\omega/2)$, although there is still visible oscillations in low frequency. The system almost reaches the thermal equilibrium, with $F(\omega)\approx \tanh (\beta\omega/2)$ at $t=30$. This matches the time scale of restoring $\mathcal{A}_{++}(\omega,0)=\mathcal{A}_{--}(\omega,0)$ and $\mathcal{A}_{+-}(\omega,0)=0$.

We further introduces $\Delta_1$ and $\Delta_2$ to quantify the strength of the Kramers' degeneracy breaking as
\begin{equation}
\begin{aligned}
\Delta_1(t)&\equiv\int dt_r~| \tilde G^>_{++}(t_r,t)- \tilde G^>_{--}(t_r,t)|,\\
\Delta_2(t)&\equiv\int dt_r~| \tilde G^>_{+-}(t_r,t)|.
\end{aligned}
\end{equation}
The numerical results are shown Fig. \ref{fig:num} (c). We find both $\Delta_1$ and $\Delta_2$ decays exponentially in the long time. As a comparison, we plot the equilibrium Green's function $|\tilde{G}^>_{++}(t,\infty)|$, the decay rate of which corresponds to the local thermalization time~\cite{maldacena2016bound}. We find their decay rates matches, with additional oscillations due to the peaks in the quasi-particle spectral $\mathcal{A}$.
  
As a comparison, we also plot results for models with symmetry $\tilde{\cal T}$ ($\tilde{\cal T}^2=1$)~\cite{ST} in Fig. \ref{fig:num} (d) and (e). We find although the diagonal components of the spectral function show similar behaviors as systems with the symmetry $\hat{\cal T}$, we have $\mathcal{A}_{+-}\neq 0$ at any time. This is consistent with the absence of the Kramers' degeneracy for systems with $\tilde{\cal T}^2=1$.

{\color{blue}\emph{Summary and Outlook.}} --  
To summarize, we find for time-reversal symmetry ${\cal T}$ satisfying ${\cal T}^2=-1$, Kramers' degeneracy in open quantum interacting fermionic systems is equivalent fo ${\cal A}_{++}(\omega)={\cal A}_{--}(\omega)$ together with ${\cal A}_{+-}(\omega)=0$. After a sudden coupling to an environment in time-reversal symmetric way, the Kramers' degeneracy experienced a breaking and restoring process. We find the revival of Kramers' degeneracy happens after the local thermalization time $t_{\rm th}$. Similar result can be obtained for time-reversal symmetry ${\cal T}$ with ${\cal T}^2=1$. But distinctively, ${\cal A}_{+-}(\omega)=0$ is not satisfied at all the time. It also means ${\cal A}_{++}(\omega)={\cal A}_{--}(\omega)$ alone can not be seen as the condition for Kramers' degeneracy.

Further, as we see, after coupling to a bath, although Kramers' degeneracy can be recovered, there is always a large portion of time Kramers' degeneracy is violated. For this reason, if we start from a pure state in Kramers' space in the initial hamiltonian, decoherence will happen and be maintained. The decoherence in the final state can be partially implied by the line shape change in the final state spectrum compared with the initial state spectrum. In this sense, we find different respects in time-reversal symmetry in open systems. If a physical result is more sensitive to phase coherence such as the quantization of the conductance in topological insulators, we argue that these results can not be protected by the revival of Kramers' degeneracy~\cite{qi81,qi82,deng2020stability}. On the other hand, like in superconductors, the pairing is more relevant to the energy degeneracy of the Kramers' pair. Therefore the superconducting phenomenon may be more stable against the environment. Even more, as we see that equilibrium or not is very important for time-reversal symmetric systems, but many transport theories are based on linear response theory which attributes transport properties as a manifestation of equilibrium correlations. We leave a careful study in these directions to future works.

{\emph{Acknowledgements.}}   P.Z. acknowledges support from the Walter Burke Institute for Theoretical Physics at Caltech. Y. C is supported by Beijing Natural Science Foundation (Z180013), and NSFC under Grant No. 11734010.


\begin{thebibliography}{40}
\bibitem [{ST()}]{ST}%
  \BibitemOpen
  \href@noop {} {}\bibinfo {note} {We consider the model with $\tilde{T}$ by
  choosing $h~=~\mu~I\otimes I+K_x \tau_x\otimes I+K_z \tau_z\otimes I+J_x
  \tau_x\otimes \sigma_y+J_y \tau_y\otimes \sigma_z+J_z \tau_z\otimes \sigma_y$
  and $\tilde h~=~\tilde \mu~I\otimes I+\tilde K_x \tau_x\otimes I+\tilde K_z
  \tau_z\otimes I+\tilde J_x \tau_x\otimes \sigma_y+\tilde J_y \tau_y\otimes
  \sigma_z+\tilde J_z \tau_z\otimes \sigma_y$.}\BibitemShut {Stop}%
\end{thebibliography}
\end{document}